\documentclass{article}
\usepackage[preprint]{neurips_2022}
\usepackage[utf8]{inputenc} 
\usepackage[T1]{fontenc}    
\usepackage{hyperref}       
\usepackage{url}            
\usepackage{booktabs}       
\usepackage{amsfonts}       
\usepackage{nicefrac}       
\usepackage{microtype}      
\usepackage{xcolor}         
\usepackage{amsmath}
\usepackage{hyperref}
\usepackage{graphicx}
\usepackage{subcaption}
\title{Accelerating OpenPangu Inference on \textbf{NPU} via Speculative Decoding}
\author{%
Yuntao Dai, {Jing Wu}, {Hang Gu},{Teng Wang} \\
School of Software Engineering \\
University of Science and Technology of China \\
}
\begin{document}
\maketitle
\begin{abstract}
To mitigate the Memory Wall bottleneck encountered by Large Language Models (LLMs) during inference on \textbf{NPU} hardware, and addressing the scarcity of native support for mainstream speculative decoding algorithms on domestic infrastructure, this study presents an end-to-end speculative inference acceleration scheme for OpenPangu-7B. 
We adapt and enhance the Medusa algorithm through the design of a lightweight Multi-Head prediction architecture and a static Tree Attention verification mechanism. 
Empirical evaluations substantiate that our method achieves a \textbf{1.35x} end-to-end speedup on \textbf{NPU} for short-sequence generation tasks. 
Furthermore, we provide an in-depth analysis of the memory bandwidth constraints affecting acceleration gains when processing long sequences, offering critical insights for future hardware-aware optimizations. 
The code is available at: \url{https://github.com/wujing215/OpenPangu7B-with-Medusa}.
\end{abstract}
\vspace{1em}
\section{Introduction}
The advent of Transformer-based Large Language Models (LLMs) \cite{vaswani2017attention, radford2019language, brown2020language} has revolutionized the landscape of artificial intelligence, facilitating unprecedented capabilities in natural language understanding and generation. 
However, the deployment of these models, which often possess billions of parameters \cite{touvron2023llama, chowdhery2023palm}, encounters a fundamental hardware bottleneck: the discrepancy between computing power and memory bandwidth \cite{mutlu2013memory}. 
Standard autoregressive decoding generates tokens sequentially; for each token generated, the entire model weights must be transferred from High Bandwidth Memory (HBM) to on-chip computing units. 
This memory-bound characteristic results in extremely low arithmetic intensity \cite{williams2009roofline}, leaving powerful matrix computation units—such as the Tensor Cores in \textbf{NPUs}—idle for the vast majority of the inference cycle. 
To mitigate this "Memory Wall," Speculative Inference has emerged as a promising paradigm. 
By leveraging lighter proxy computations to draft multiple candidate tokens which are then verified in parallel by the target model, speculative decoding effectively converts memory-bound sequential operations into compute-bound parallel operations \cite{leviathan2023fast, chen2023accelerating, kim2023speculative}. 
Among these approaches, Medusa \cite{cai2024medusa} represents the state-of-the-art by appending lightweight Multi-Layer Perceptron (MLP) heads to the frozen backbone, thereby achieving block-wise generation without the complexity of maintaining a separate draft model. 
Despite the efficacy of Medusa on NVIDIA GPU platforms, transplanting this paradigm to the emerging domestic \textbf{NPU} hardware ecosystem presents non-trivial challenges. 
First, there is a distinct software-hardware misalignment. 
Current implementations heavily rely on dynamic control flow and CUDA-specific optimizations \cite{pope2023efficiently} which are incompatible with the \textbf{NPU} software stack. 
Second, and more critically, there exists a conflict between the algorithmic nature of speculative decoding and the execution model of the NPU. 
The Medusa algorithm inherently requires dynamic tree construction and verification based on real-time prediction probabilities \cite{miao2023specinfer}. 
In contrast, the \textbf{NPU} architecture is optimized for Static Graph execution \cite{liao2021ascend}, where memory addresses and computation graphs are fixed prior to runtime to maximize throughput. 
Implementing dynamic tree attention logic directly typically triggers frequent graph recompilation and severe synchronization overheads \cite{aminabadi2022deepspeed}, often negating any potential speedup. 
In response to these challenges, this paper presents an end-to-end acceleration scheme for OpenPangu-7B \cite{chen2025pangu} that bridges the gap between advanced speculative algorithms and domestic hardware characteristics. 
Our primary contributions are summarized as follows: 
\begin{itemize}
\item We establish the first comprehensive implementation of the Medusa speculative decoding framework within the \textbf{NPU} PyTorch ecosystem, systematically resolving operator-level incompatibilities in the software stack. 
\item We propose a hardware-friendly static tree construction mechanism paired with a zero-copy retrieval strategy, reconciling the conflict between dynamic speculative verification and the NPU's static execution model. 
\item We conduct extensive empirical benchmarks on the \textbf{NPU} platform, demonstrating a 1.35$\times$ end-to-end speedup while providing a critical analysis of memory bandwidth constraints \cite{dao2022flashattention, dao2023flashattention2}. 
\end{itemize}
\section{Background}
\subsection{The Memory Wall in Autoregressive Generation}
The performance of LLM inference is governed by the operational intensity of the workload relative to the hardware's operational intensity capacity \cite{williams2009roofline}. 
In the standard autoregressive decoding phase, the generation of a single token $x_t$ conditioned on the context $x_{<t}$ requires a full forward pass of the model parameters $\theta$. 
The arithmetic intensity (OPS/Byte) of this operation is bounded by the batch size. 
For latency-sensitive applications where the batch size is small, the arithmetic intensity drops significantly below the machine balance point \cite{pope2023efficiently, kwon2023efficient}. 
\textbf{NPU } processors are designed with massive parallel compute units (Cube Cores) optimized for dense matrix multiplications. 
However, the latency is dominated by the memory access time required to fetch weights from HBM. 
This phenomenon implies that simply increasing compute capability (FLOPS) yields diminishing returns for inference latency unless the algorithmic paradigm is fundamentally altered \cite{shazeer2019fast}. 
\subsection{Evolution of Speculative Decoding Paradigms}
Speculative decoding aims to decouple memory access frequency from the token generation rate. 
Early works \cite{leviathan2023fast, chen2023accelerating} introduced the "Draft Model" concept—a smaller auxiliary model used to generate candidate sequences. 
While theoretically sound, the Draft Model approach faces practical deployment hurdles: managing two distinct models in memory \cite{spector2023accelerating}, complicates serving infrastructure, and necessitates tokenizer alignment. 
To address these, "Model-Free" approaches like Lookahead Decoding \cite{fu2023break} were proposed, utilizing n-gram matching. 
Furthermore, "Self-Speculation" paradigms, exemplified by Medusa \cite{cai2024medusa} and Eagle \cite{li2024eagle}, integrate drafting into the main model. 
By adding auxiliary heads, Medusa achieves high acceptance rates with negligible overhead. 
However, its tree-based verification remains a bottleneck on static-graph-oriented architectures \cite{miao2023specinfer}. 
\subsection{Architectural Constraints of \textbf{NPU}}
\begin{figure}[htbp]
\centering
\begin{subfigure}[b]{0.48\textwidth}
\centering
\includegraphics[width=\textwidth]{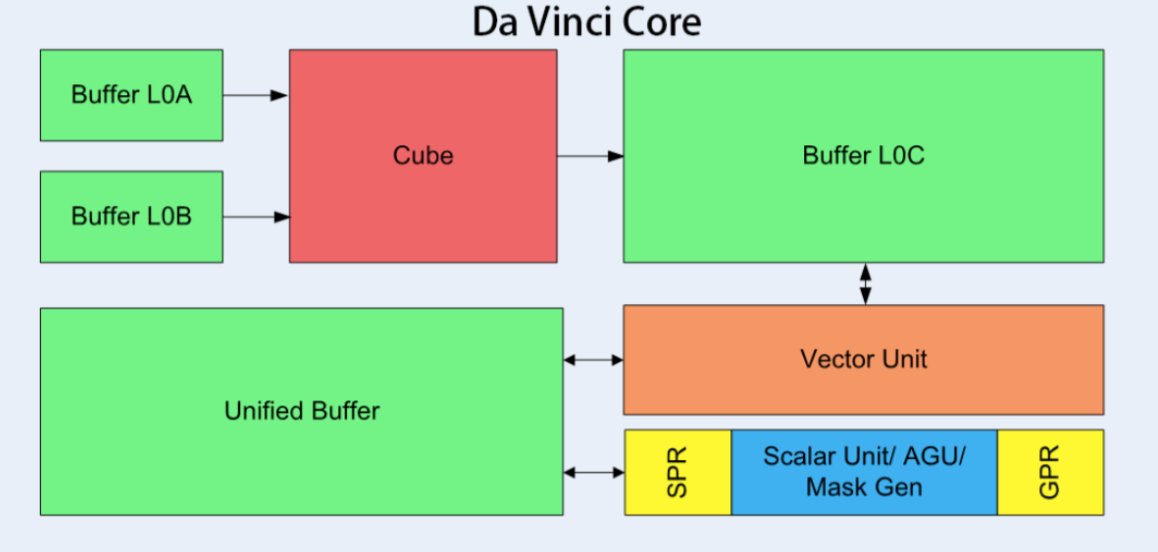} 
\caption{The Core Architecture}
\label{fig:davinci_core}
\end{subfigure}
\hfill
\begin{subfigure}[b]{0.48\textwidth}
\centering
\includegraphics[width=\textwidth]{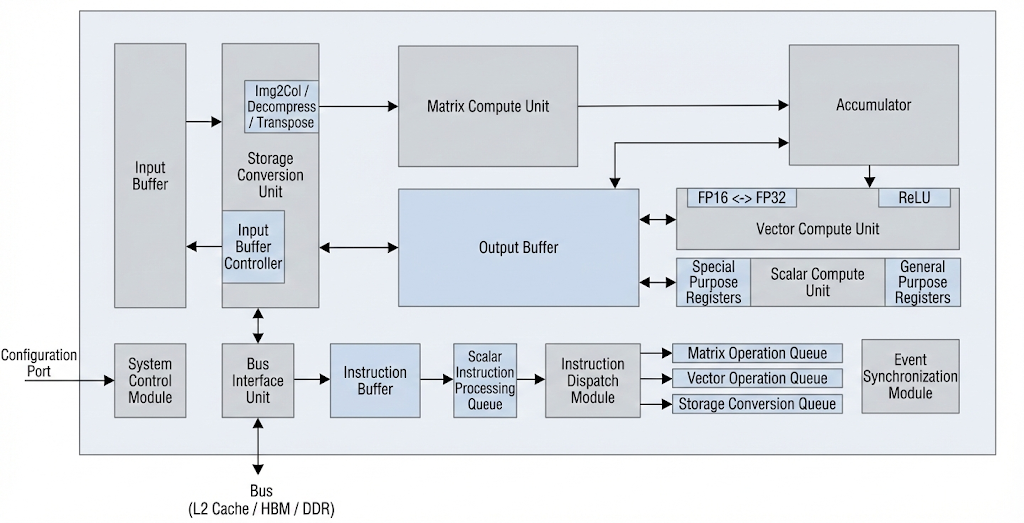} 
\caption{Data Flow and Compute Units}
\label{fig:ascend_flow}
\end{subfigure}
\caption{\textbf{Micro-architecture of the \textbf{NPU} Processor.} (a) The Core separates Matrix (Cube) and Vector operations, necessitating distinct buffer management (L0A/L0B/L0C). (b) The detailed data flow shows that instruction dispatch and memory movement are highly pipelined. This hardware design heavily favors \textbf{Static Shape} execution, where data movement paths are determined at compile-time, contrasting with the dynamic scheduling flexibility of GPUs.}
\label{fig:davinci}
\end{figure}

Unlike GPUs which offer flexible thread scheduling, the \textbf{NPU} Processor emphasizes high throughput via a static execution model \cite{liao2021ascend}. 
As illustrated in Figure \ref{fig:davinci}, the underlying architecture relies on a specialized hierarchy of computing units (Cube and Vector cores) and explicit memory management (L0/L1 Buffers). 
The \textbf{NPU} software stack performs optimal operator fusion and memory allocation during an offline compilation phase. 
This \textbf{Static Shape} mechanism assumes invariant input tensor shapes and computation graph topology. 
Speculative decoding algorithms involve conditional branches: the number of tokens accepted and the shape of the attention mask change at every step. 
On \textbf{NPUs}, such dynamism triggers expensive Just-In-Time (JIT) recompilation or CPU fallback, introducing prohibitive latency. 
The \textbf{NPU} software stack performs optimal operator fusion and memory allocation during an offline compilation phase based on this architecture. 
This mechanism, known as \textbf{Static Shape}, assumes that the input tensor shapes and the computation graph topology remain invariant during execution. 
Speculative decoding algorithms, by definition, involve conditional branches: the number of tokens accepted and the shape of the attention mask effectively change at every step based on verification results. 
On an NVIDIA GPU, this is handled via dynamic kernel launches. 
On an \textbf{NPU}, however, such dynamism forces the runtime to either fall back to scalar execution on the CPU or trigger expensive Just-In-Time (JIT) recompilation, both of which introduce prohibitive latency penalties. 
Therefore, enabling efficient speculative decoding on \textbf{NPU} requires a paradigm shift from dynamic logic to static, tensor-based formulations, which constitutes the core technical contribution of this work. 
\section{Methodology}
\subsection{Multi-Head Prediction Architecture}
To empower the OpenPangu-7B backbone with future prediction capabilities while maintaining architectural consistency, we employ a non-invasive adaptation strategy. 
Specifically, we append $K$ parallel decoding heads, designated as Medusa Heads, immediately following the final Transformer layer of the backbone model. 
Architecturally, each head is instantiated as a Multi-Layer Perceptron (MLP) equipped with residual connections, designed to project the hidden states from the current timestamp $t$ directly into the probability distributions of future tokens. 
The training of these auxiliary heads is decoupled from the frozen backbone to preserve the original model's generation quality. 
The optimization objective is formulated as the weighted sum of cross-entropy losses across all heads. 
Let $h_t$ denote the last hidden state at step $t$, and $p_k(h_t)$ represent the probability distribution predicted by the $k$-th head for the token at $t+k+1$. 
The total loss function $\mathcal{L}$ is defined as: 
\begin{equation}
\mathcal{L} = \sum_{k=1}^{K} \lambda_k \cdot \text{CE}(p_k(h_t), x_{t+k+1})
\end{equation}
where $\lambda_k$ serves as a decay coefficient. 
This weighting scheme reflects the increasing uncertainty associated with predicting more distant tokens, thereby stabilizing the convergence of the multi-head training process. 
\begin{figure}[htbp]
\centering
\includegraphics[width=0.95\textwidth]{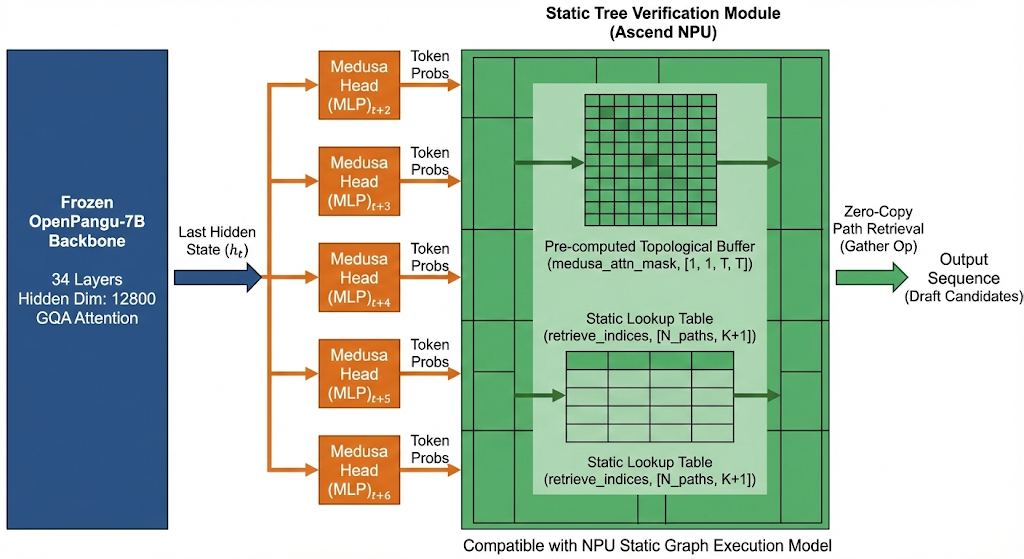} 
\caption{\textbf{Overview of the OpenPangu-with-Medusa architecture adapted for \textbf{NPU}.} The system features (a) a frozen OpenPangu backbone with lightweight MLP heads for block-wise token prediction, and (b) a \textit{Static Tree Verification} module. The latter utilizes pre-computed topological buffers (\texttt{medusa\_attn\_mask}) to enable zero-copy path retrieval, ensuring compatibility with the NPU's static graph execution model.}
\label{fig:arch}
\end{figure}
\subsection{Static Implementation for NPU}
The primary obstacle to deploying speculative decoding on \textbf{NPUs} lies in the hardware's preference for Static Shape execution. 
Dynamic variations in tree topology typically trigger frequent graph recompilation, neutralizing potential speedups. 
To address this, we propose a fully static implementation paradigm that eliminates dynamic control flows. 
\paragraph{Tensorization of Tree Topology.}
Instead of dynamically constructing the verification tree at runtime, we pre-calculate the topological structure offline and map it to static tensors. 
We define a static visibility matrix, \texttt{medusa\_attn\_mask}, with dimensions $[1, 1, T, T]$, which encodes the parent-child dependencies of all candidate paths within the tree. 
Simultaneously, we construct a \texttt{tree\_indices} tensor that creates a deterministic mapping between the flattened tree nodes and their corresponding positions in the KV-cache. 
By loading these invariant tensors into the NPU memory during the initialization phase, we ensure that the computational graph remains constant regardless of the speculative verification outcome, allowing the compiler to perform aggressive operator fusion. 
\paragraph{Zero-Copy Retrieval Strategy.}
Traditional implementations often rely on CPU-side pointer manipulation to reconstruct the accepted token sequence, a process that incurs significant synchronization overhead and interrupts the NPU execution pipeline. 
We introduce a Zero-Copy Retrieval mechanism to circumvent this bottleneck. 
We construct a static lookup table, \texttt{retrieve\_indices}, of size $[N_{paths}, K+1]$, which stores the exact memory offsets for every possible candidate path. 
During the verification phase, once the optimal path index is identified via the acceptance criteria, the model utilizes the on-chip \texttt{Gather} operator to extract the final token sequence directly from the candidate buffer using the lookup table. 
This design retains all data movements within the high-bandwidth NPU memory, achieving zero-copy path reconstruction and maximizing the utilization of the static execution graph. 
\section{Experiments \& Analysis}
\subsection{Experimental Setup}
\textbf{Hardware Environment.} The experimental evaluation was conducted on a high-performance computing cluster equipped with \textbf{NPU } processors (running firmware and hosted on Kunpeng 920 CPUs). 
To provide a cross-architecture baseline, we also performed comparative benchmarks on an NVIDIA RTX A6000 GPU (Ampere Architecture) utilizing CUDA 11.8. 
\textbf{Subject Model: OpenPangu-7B.} The subject model for all experiments is \textbf{openPangu-Embedded-7B-V1.1}, a high-efficiency LLM trained from scratch on the \textbf{NPU} ecosystem \cite{chen2025pangu}. 
This model is pre-trained on 25 trillion tokens and possesses a unique capability for "fast and slow thinking" adaptation. 
Unlike generic Llama-based architectures, OpenPangu-7B is specifically optimized for domestic hardware characteristics. 
The detailed architectural parameters of the model are summarized in Table \ref{tab:openpangu_specs}. 
\begin{table}[htbp]
\centering
\caption{\textbf{Architectural Specifications of openPangu-Embedded-7B-V1.1.} (Non-embedding parameters account for approximately 7B).}
\label{tab:openpangu_specs}
\small
\begin{tabular}{ll}
\hline
\textbf{Attribute} & \textbf{Configuration} \\ \hline
Architecture & Dense \\
Parameters (Non-Embedding) & 7B \\
Number of Layers & 34 \\
Hidden Dimension & 12,800 \\
Attention Mechanism & Grouped Query Attention (GQA) \\
Number of Attention Heads & 32 for Q, 8 for KV \\
Vocabulary Size & 153k \\
Context Length (Natively) & 32k \\
Pretraining Tokens & 25T \\ \hline
\end{tabular}
\end{table}
\textbf{Implementation Details.} For the training of the auxiliary Medusa Heads, we employed the AdamW optimizer with a learning rate of $1e-3$ and a global batch size of 64. 
All inference tests utilize Float16 precision. 
\textbf{Evaluation Metrics.} To quantitatively assess the efficiency of the proposed scheme, we define three primary metrics. 
First, the \textbf{End-to-End Speedup} represents the wall-clock time improvement of our method compared to standard autoregressive decoding. 
Second, the \textbf{Accept Rate (AC)} measures the average number of tokens verified and accepted per decoding step. 
Finally, the \textbf{Overhead Ratio} quantifies the additional computational cost introduced by the speculative sampling and tree verification processes. 
The relationship between these metrics is formalized as follows: 
\begin{equation}
\text{Speedup} = \frac{\text{AC}}{\text{Overhead}}
\end{equation}
\begin{equation}
\text{Overhead} = \frac{\text{Time}_{speculative}}{\text{Time}_{autoregressive}}
\end{equation}
Where $\text{Time}_{speculative}$ denotes the latency of a single Medusa decoding step, and $\text{Time}_{autoregressive}$ denotes the latency of a standard single-token generation step. 
\subsection{Impact of Self-Distillation Data Scale}
The alignment between the prediction heads and the backbone model is critical for high acceptance rates. 
To investigate the data dependency of Medusa Heads, we trained multiple configurations using both a public dataset (ShareGPT) and a custom Self-Distillation dataset. 
The self-distillation dataset was constructed by prompting the OpenPangu-7B backbone with queries from ShareGPT and collecting its output logits as soft labels. 
This process ensures that the heads learn the specific probability distribution of the host model. 
We scaled the training data from 2k to 50k samples and evaluated the Top-1 accuracy of the predicted tokens. 
Table \ref{tab:distillation} summarizes the performance across different training configurations. 
Initial experiments relying solely on the public ShareGPT dataset (2k samples) yielded a suboptimal Top-1 accuracy of 62.40\% for the first head, indicating a distribution shift between the generic chat data and the model's specific behavior. 
\begin{table}[htbp]
\centering
\caption{Performance Comparison of Medusa Heads with Different Training Datasets}
\label{tab:distillation}
\resizebox{\textwidth}{!}{%
\begin{tabular}{ccccccc}
\toprule
\textbf{Dataset Size} & \textbf{Number of} & \textbf{Train with} & \textbf{Reserve} & \textbf{Top-1 Acc} & \textbf{Top-1 Acc} & \textbf{Training Time} \\
\textbf{(Samples)} & \textbf{Heads} & \textbf{Self-Distillation?} & \textbf{Special Tokens?} & \textbf{(Head 1)} & \textbf{(Head 2)} & \textbf{(Hours)} \\ \midrule
2k & 3 & No & No & 62.40\% & 42.00\% & 7.45 \\
10k & 5 & Yes & No & 67.80\% & 49.30\% & 63.00 \\
\textbf{50k} & \textbf{5} & \textbf{Yes} & \textbf{Yes} & \textbf{74.60\%} & \textbf{54.10\%} & \textbf{65.70} \\ \bottomrule
\end{tabular}%
}
\end{table}
Introducing self-distillation (10k samples) improved the accuracy to 67.80\%, yet the alignment remained imperfect. 
A deeper analysis revealed that the initial distillation process inadvertently filtered out OpenPangu's special control tokens (e.g., tokens marking internal "thinking" states or start/end boundaries). 
Consequently, the heads failed to learn the structural formatting norms of the backbone. 
In the final configuration (50k samples), we explicitly preserved these special tokens during the data generation and training phases. 
This optimization proved decisive, boosting the Top-1 accuracy of Head 1 to \textbf{74.60\%} and Head 2 to \textbf{54.10\%}. 
This result underscores that for specialized LLMs like OpenPangu, the prediction heads must be trained to mimic not only the semantic content but also the exact token-level formatting quirks of the host model. 
\begin{figure}[t]
\centering
\includegraphics[width=0.85\textwidth]{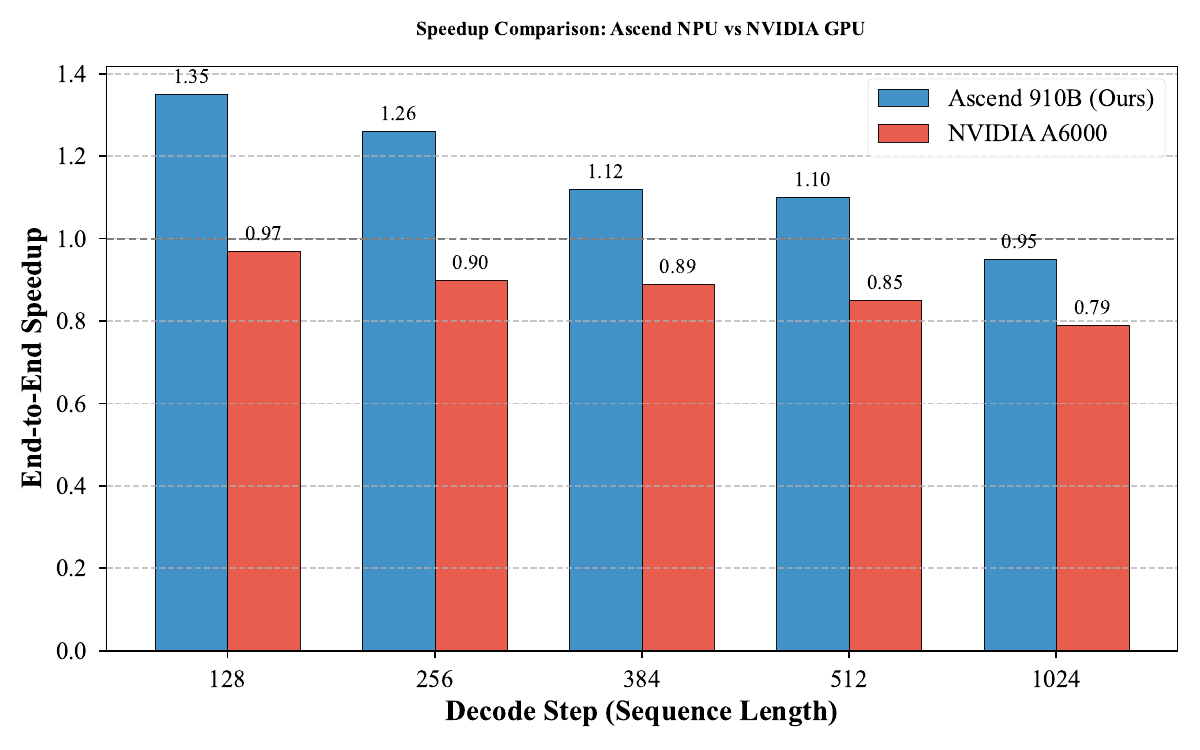}
\caption{\textbf{End-to-End Speedup Comparison (\textbf{NPU } vs. NVIDIA A6000).} The proposed method achieves a peak speedup of \textbf{1.35$\times$} on \textbf{NPU} for short sequences ($L=128$), significantly outperforming the unoptimized GPU baseline. However, the speedup on NPU exhibits a downward trend as sequence length increases, eventually crossing the break-even point at $L=1024$.}
\label{fig:speedup}
\end{figure}
\begin{figure}[h]
\centering
\includegraphics[width=0.85\textwidth]{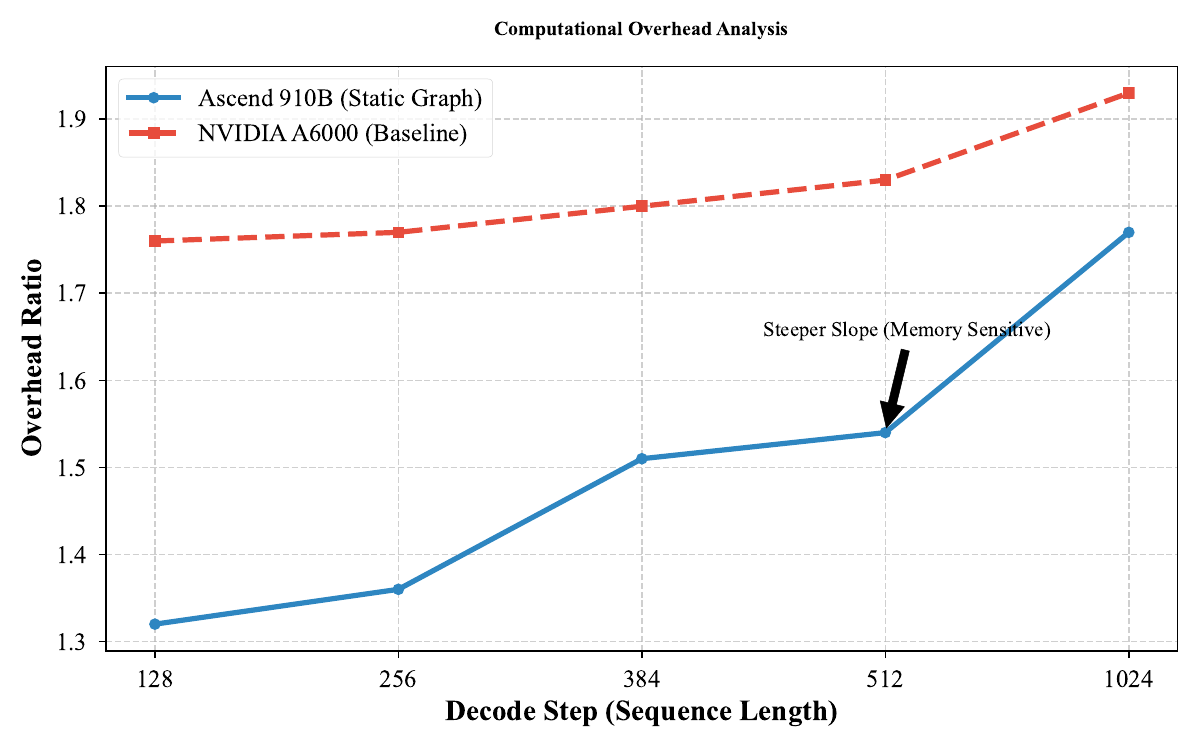}
\caption{\textbf{Computational Overhead Analysis.} The graph illustrates the ratio of speculative decoding time to standard autoregressive time. The \textbf{NPU} (blue line) shows a steeper slope compared to the GPU (red dashed line), indicating higher sensitivity to memory access patterns. This non-linear growth in overhead is the primary factor limiting speedup in long-context scenarios on the NPU.}
\label{fig:overhead}
\end{figure}
\subsection{In-depth Performance Analysis on \textbf{NPU}}
We further analyzed the dynamic performance characteristics of the system across varying sequence lengths. 
On the \textbf{NPU } platform, our method demonstrates a robust speedup of \textbf{1.35$\times$} for short sequences (Decode Step 128). 
This performance advantage is attributed to the efficient parallelization of the matrix operations within the static graph, where the overhead ratio is kept low (1.32). 
However, a distinct trend emerges as the sequence length extends. 
While the Accept Rate remains relatively stable (dropping only slightly from 1.78 to 1.65), the computational Overhead rises non-linearly. 
As illustrated in Figure \ref{fig:overhead}, the overhead on \textbf{NPU} reaches 1.77 at a sequence length of 1024. 
This phenomenon highlights the Memory Wall constraint inherent in the hardware. 
As the KV-cache grows linearly, the attention mechanism's memory access pattern becomes increasingly sparse and bandwidth-intensive. 
The \textbf{NPU}, while powerful in computation, exhibits higher sensitivity to these memory-bound operations compared to the GPU baseline. 
Consequently, the speedup gain diminishes in long-context scenarios, suggesting that future optimizations must prioritize memory access patterns, such as KV-cache compression or tiling strategies, over pure compute optimization. 
\section{Conclusion}
This paper presented an end-to-end acceleration scheme for OpenPangu-7B on \textbf{NPU} hardware. 
By introducing \textbf{Static Tree Construction} and \textbf{Zero-Copy Retrieval}, we overcame the incompatibility between dynamic speculative decoding and NPU static graph execution. 
Results show a \textbf{1.35x speedup} on \textbf{NPU }. 
Future work will focus on operator fusion and memory optimization for long-context scenarios. 
\bibliographystyle{plainnat}
\bibliography{ref}
\end{document}